\documentclass[a4paper]{article}
\usepackage{epsfig,amsfonts,amssymb,setspace,multicol,textcomp,xcolor}
\usepackage[T1]{fontenc}
\usepackage{amsmath}
\begin{document}
\fontsize{11}{11} \selectfont
\title{\bf Using THELI pipeline in order to reduce Abell 226\\ multi-band optical images}
\author{\textsl{R.~Joveini$^{1,2}$\thanks{joveini@physics.sharif.edu}, S.~Sotoudeh$^{1,2}$, A.~Roozrokh$^{2,3}$, M.~Taheri$^{1}$}}
\date{}
\maketitle
\begin{center} 
{\small $^{1}$Physics Department, Sharif University of Technology, Azadi  Ave., Tehran, Iran\\
$^{2}$School of Astronomy, Institute for Research in Fundamental Sciences, Opposite Araj, Artesh Highway, Tehran, Iran\\
$^{3}$Physics and Astronomy Department, University of California Riverside, 900 University Ave., Riverside, CA 92521, USA\\}
\end{center}
\begin{abstract}
In this paper we review {\tt THELI} (Erben \& Schrimer, 2005), an image processing pipeline developed to reduce multi-pointing optical images taken by mosaic CCD cameras. This pipeline works on raw images by removing several instrumental contaminations, implementing photometric calibration and astrometric alignment, and constructing a deep co-added mosaic image complemented by a weight map. We demonstrate the procedure of reducing NGC\,3923 images from raw data to the final results. We also demonstrate the quality of our data reduction strategy using mag-count and mag-error in mag plots. Emphasis is mainly placed on photometric calibration which is of great interest to us due to our scientific case. Based on the cross-association of the extracted catalogue against a reference catalogue of stellar magnitudes, zero-point calibration is performed. Our data reduction strategy and the method employed for cross-correlating large catalogues is also presented.\\[1ex]
\end{abstract}

\section*{\sc introduction}
\indent \indent In astronomy, raw CCD images from telescopes cannot be directly used for scientific purposes. Firstly, several instrumental effects must be removed. Generally speaking, data reduction is the transformation of raw data into a form suitable for analysis. The following processes are required to achieve this goal: 
\begin{itemize}
\item removal of instrumental signatures, such as bias offset, dark currents, field curvature (caused by projecting the spherical sky onto a flat CCD plane), and fringe patterns;
\item masking of unwanted signals, such as cosmic rays, stellar halos and satellite tracks;
\item photometric and astrometric calibration;
\item co-addition of individual frames. 
\end{itemize}
The {\tt THELI} (Transforming HEavenly Light into Image)\footnote{to be obtained from {\tt http://astro.uni-bonn.de/\texttildelow{}mischa/theli.html}} pipeline was initially developed for WFI cameras (it consists of 8 CCD chips of 2k$\times$4k each)\footnote{for more information, see: {\tt http://www.eso.org/sci/facilities/lasilla/instruments/wfi/index.html}} mounted on the ESO 2.2~meter Max Planck telescope. It has a modular design allowing it to be adapted to other single- or multi-chip cameras with considerable ease.

The primary goal of developing the {\tt THELI} pipeline was to reduce weak lensing data; therefore, more emphasis was laid on the accurate alignment of galaxies for each exposure (precise astrometry, rather than photometry), acquiring the highest possible resolution and accurate noise mapping.

Through the course of the data reduction process the {\tt THELI} pipeline uses well-tested astronomical software. This allows for easy exchange whenever a new algorithm or a better implementation becomes available. The preliminary tools in the {\tt THELI} pipeline are the {\tt LDAC}\footnote{Leiden Data Analysis Center, available at: {\tt ftp://ftp.strw.leidenuniv.nl/pub/ldac/software}} software \cite{deul99}, the {\tt TERAPIX} software \cite{bertin02}, {\tt Eclipse} and {\tt qfits} tools\footnote{available at: {\tt http://www.eso.org/projects/aot/eclipse}} and {\tt IMCAT} utilities\footnote{{\tt http://www.ifa.hawaii.edu/\texttildelow{}kaiser/imcat/}}.

Next, the processing a set of optical images in the ESO bands B, R, and I is described. Our initial data set consisted of $30$~Gigabytes of raw exposures: {\tt SCIENCE} frames of 7 different galaxy groups and calibration frames taken in 3 successive nights with WFI at ESO  2.2m\footnote{ESO Programme ID of the observations: 077.A-0747(A) on April 4, 2006}. Prior to the reduction procedure all objects are treated in the same manner, i.\,e. we suppose the same calibration frames for all of the {\tt SCIENCE} frames.

\section*{\sc pre-reduction}
\indent \indent Algorithms in the pipeline which remove instrument effects (such as bad CCD pixels and fringe patterns) and which can be considered constant during the period of our observation (3 nights), are described further. 

Firstly, each file's header is updated with the necessary keywords for the pipeline. The image is then divided into the number of chips in the CCD (which constitutes 8 CCD chips in a WFI mosaic). From this step on, the pipeline works on individual chips rather than whole images, thereby enhancing the speed and enabling us to do multi-chip processing on multi CPUs.

\subsection*{\sc bias and dark frame}
\indent \indent In the first step, following a preliminary inspection of all exposures of a given type (bias, flat field), the pipeline identifies a typical count level and rejects outliers, e.\,g. {\tt DARKs} with too-short exposure time or over-saturated {\tt SKYFLATs}. The acceptable range is defined by the user. Since ESO WFI is constantly cooled to a stable temperature of 167\,K \cite{handbook} and dark current is negligible, many observers do not take {\tt DARK} frames. We will describe their usage when appropriate,  however it should be noted that we have not had or processed DARK frames. 
 
Each frame is overscan-corrected (OC) and trimmed: overscan region is trimmed off the image. Master {\tt BIAS} and master {\tt DARK} 
frames are created by (median or arithmetic) averaging of individual {\tt BIAS} and {\tt DARK} frames. The pipeline uses master {\tt DARK} to identify bad pixels. 

\subsection*{\sc flat-field pattern}
\indent \indent In general, the sensitivity and the illumination of a CCD is not homogeneous. Sensitivity can change from pixel to pixel, while the shift in illumination is considerable only at larger scales of distance. To make surface brightness homogeneous, we utilise a flat-field pattern. Due to the dependence of the flux of radiation of astronomical objects and the sky background on the filter used, the tasks described below need to be carried out separately for each filter.

\subsection*{\sc manual inspection, mask creation}
\indent \indent An 8$\times$8 binned mosaic of corrected {\tt SCIENCE} frames is created to check the run process. At this stage, we looked at binned mosaics to check if pre-reduction was performed correctly. Using {\tt DS9} software, we manually omitted out-of-focus data, and masked extended defects like satellite tracks and bright star reflections.

\subsection*{\sc creating weight frames}
\indent \indent The pipeline creates a weight frame for each {\tt SCIENCE} frame. Hot/cold pixels are detected by studying the master {\tt DARK}, saturated pixels are identified by thresholding the {\tt SCIENCE} frames; cosmic rays are discovered by {\tt SExtractor} \cite{bertin96} in connection with {\tt EyE} \cite{bertin01}, and manual masks are added using the {\tt LDAC} utilities.

Global {\tt Weight} and {\tt Flag} frames are made for each CCD chip. Flag frames are integer {\tt FITS} in which 0 denotes good pixels and every other value denotes a certain defect. The values are used when producing weights. Global {\tt Weights} contain information about bad pixels of all images from that chip.

\section*{\sc reduction}
\indent \indent We group {\tt SCIENCE} frames into sets, depending on the objects they contain. Our pointings are: HCG\,48, HCG\,62, HCG\,67, NGC\,3557, NGC\,3923, NGC\,4697 and RX-J2114.3-6800. We run the reduction process on each pointing and subsequently add the reduced 
individual frames.

\subsection*{\sc astrometery}
\indent \indent The initial step in astrometery is to detect high S/N objects using {\tt SExtractor} and to generate a catalogue of non-saturated stars. By comparing this catalogue to the USNO-B1 catalogue, a zero-order, single shift astrometric solution is calculated for each image. The CCDs in multi-chip cameras can be rotated unintentionally with respect to each other. In addition, due to a large field of view, the sky must be considered to be a spherically curved surface.

The next step is to estimate third-order polynomials for the astrometric solution for each chip, using the {\tt SCamp} package \cite{bertin06} developed by {\tt TERAPIX}.

\subsection*{\sc photometry} 
\indent \indent Exposures are taken under varying conditions: parameters such as airmass and background radiation differ from one night of observation to another. Therefore, during the first part of photometric calibration, images should be calibrated relatively. This is done by the {\tt LDAC relphotpm} program. {\tt Relphotom} takes tables of overlapping exposures as input and calculates the mean deviation of magnitudes:
\begin{equation}
M_{k,j}=\frac{\sum\limits_{i}\left(\sigma^{2}_{K}+\sigma^{2}_{J}\right)^{-1}\left(M_{K_{i}}-M_{J_{i}}\right)}{
\sum\limits_{i}\left(\sigma^{2}_{K}+\sigma^{2}_{J}\right)^{-1}},
\end{equation}
where $K$ and $J$ $(K \neq J)$ are fields, $i$ denotes objects present in both exposures, and $\sigma$  are measurement errors of the magnitude, and finds the relative zero point magnitude by $\chi^{2}$ minimisation:
\begin{equation}
\chi^{2}=\sum\limits_{k,j}^{N}\left[M_{k,j}-\left(ZP_{k}-ZP_{j}\right)\right]^2.
\end{equation}
In this case we neglect variant photometric conditions which affect each spectral type in a different way.

\section*{\sc co-addition}
\subsection*{\sc preparation for co-adding}
\indent\indent Sky background is calculated using {\tt SExtractor BACKGROUND} check image for every large-object-subtracted frame, and is subtracted from all {\tt SCIENCE} images. A re-sampled image is made using the astrometric solution polynomials calculated in astrometry step and stored in each chip's header.

\subsection*{\sc co-addition of individual frames} 
\indent\indent {\tt THELI} offers two tools for image co-adding: {\tt SWarp} and EIS {\tt Drizzle}. Our tool of choice is the former. Using all input images (which are related to a single output pixel) in our sample, {\tt SWarp} calculates the final results using the weighted mean method.

A {\tt WEIGHT} image is also produced, as well as a {\tt FLAG} image. {\tt WEIGHT} image plays a key role in source detection of the final image \cite{erben05}, for a comparison between weighted and un-weighted source detection.

\section*{\sc our method of catalogue \\cross association}
\indent \indent As described above, we need to match object information from different channels. Generally there are two cases when it is necessary to consult two or more catalogues:
\begin{enumerate}
 \item When there is given a catalogue in a particular filter and it is desired to cross-match it with a reference catalogue, like a Standard Star Magnitude catalogue.
  \item When there are multiple observations of a source in different filters. In this case, one needs to deal with multiple channels (in our case: B, R and I).
 \end{enumerate}
Since {\tt SExtractor}'s detection is not perfect in terms of source coordinates, one has to take one catalogue as a reference catalogue and treat the others as search catalogues; therefore, for each object in a certain channel, information from the other channels will be found and added, as to make a ``matched and merged'' catalogue.

\section*{\sc results and qualifications}
\begin{figure}
\centering
\begin{minipage}[t]{.48\linewidth}
\centering
\epsfig{file = 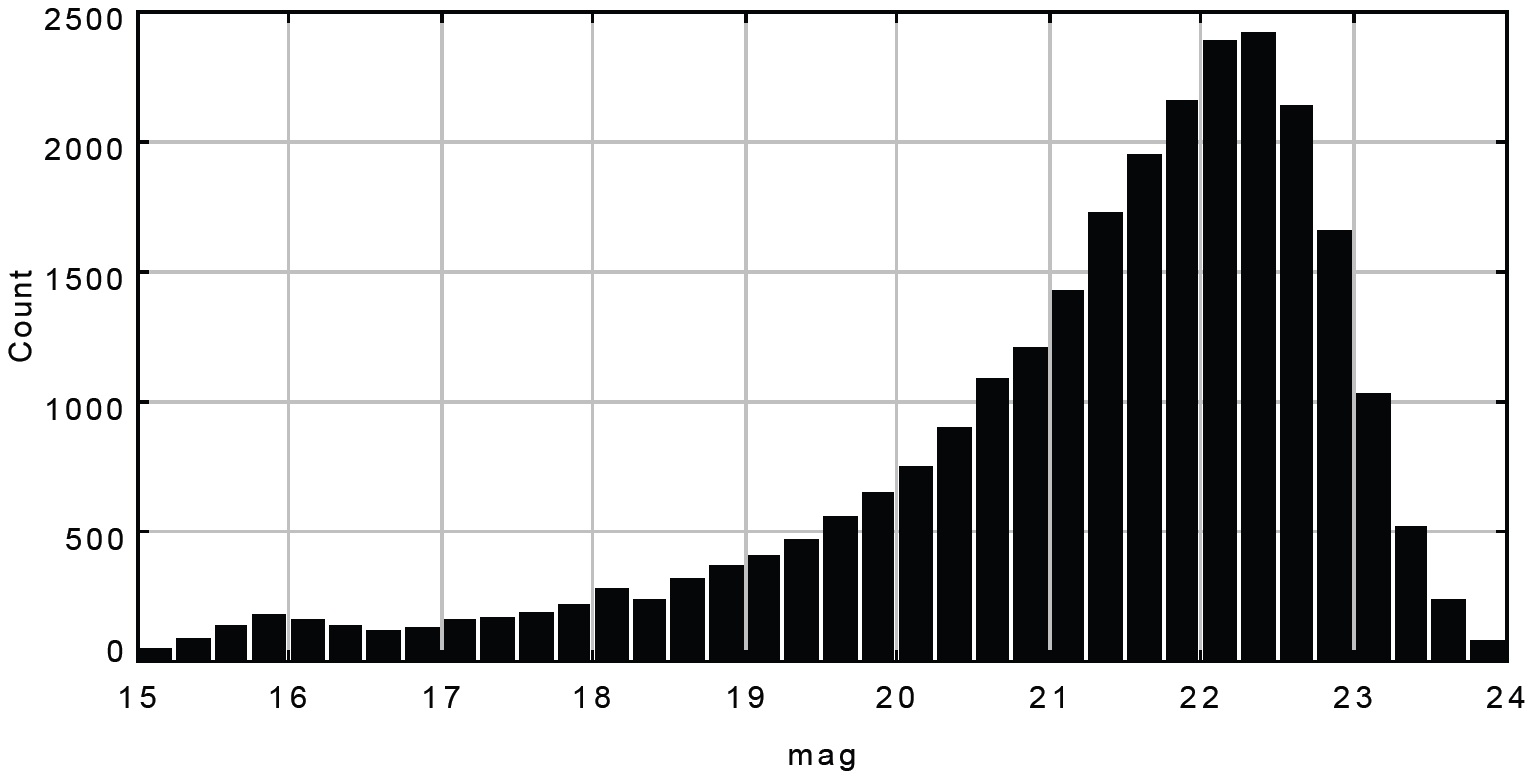,width = .99\linewidth}
\caption{Qualification plot for NGC\,3923 image reduction: galaxy count histogram for NGC\,3923 field.}\label{fig1a}
\end{minipage}
\hfill
\begin{minipage}[t]{.48\linewidth}
\centering
\epsfig{file = 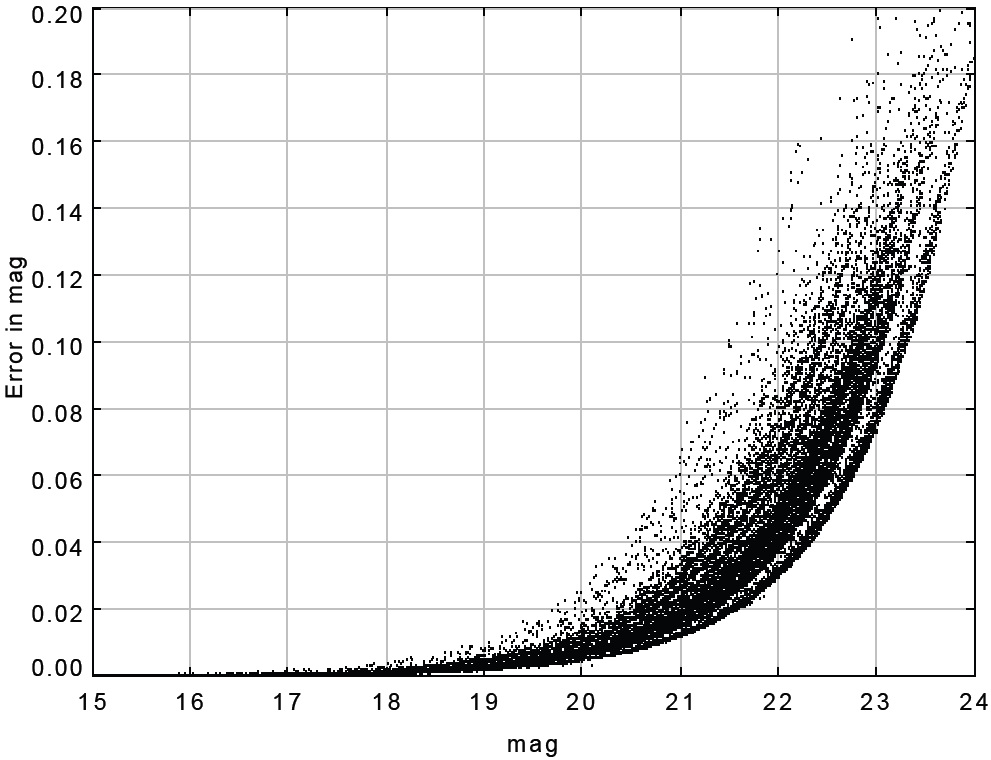,width = .99\linewidth}
\caption{Qualification plot for NGC3\,923 image reduction: galaxy magnitude error scatter plot for NGC\,3923 field.}\label{fig1b}
\end{minipage}
\end{figure}

\indent\indent With the final {\tt COADD} images along with their {\tt WEIGHTs} in hand in B, R and I filters, one would produce several plots to show the qualification of data reduction. Throughout this, two of them are very important: mag-count histogram and mag-error in mag scatter plot. As shown in Fig.\,\ref{fig1a}  we have plotted the mag-count histogram for NGC\,3923 field of galaxies (note that stars have been separated from galaxies using {\tt SExtractor}'s {\tt CLASS\_STAR} parameter larger than 0.95 as a criteria. Seven outliers have been removed, since they were only round-shaped galaxies). As seen, the count drops from magnitude 22.5; this is called completeness limit. It shows that our data obtained with this instrument is trustworthy up to this completeness limit (note that the count should be raised with respect to the fainter objects).

The other plot, which is important in astronomical data reduction, is the mag-error in mag scatter plot as we showed in Fig.\,\ref{fig1b}. This plot reveals which data is trustworthy and which data is not based on our scientific goal. It demonstrates that for objects with magnitude around 22.5 the error in magnitudes increase very fast; therefore objects fainter than 22.5 are not suitable for this kind of investigation. 

As a fundamental parameter in astronomy photometric data, seeing\footnote{calculated using FWHM in an area around image centre} associated with the individual pointings (in R band) ranged from 0.75 to 1.1~arcsec, and the average seeing of the co-added image has been measured to be 0.89~arcsec.

\section*{\sc acknowledgement}
\indent \indent The authors acknowledge IPM for providing computational facilities and also appreciate Reza Mansouri, Habib G. Khosroshahi and Alireza Molaeinezhad for useful discussions and guidance.

\end{document}